\documentclass[12pt]{iopart}

\usepackage{iopams}
\usepackage{amsmath2}
\usepackage{graphicx}
\usepackage{color}
\usepackage{amssymb}

\newcommand{\tn}{\textnormal}

\begin{document}

\title[Hubbard-to-Heisenberg crossover of Drude weights at low temperatures]{Hubbard-to-Heisenberg crossover (and efficient computation) of Drude weights at low temperatures}

\author{C. Karrasch}

\address{Dahlem Center for Complex Quantum Systems and Fachbereich Physik, Freie Universit\"at Berlin, 14195 Berlin, Germany}

\ead{c.karrasch@fu-berlin.de}

\begin{abstract}

We illustrate how finite-temperature charge and thermal Drude weights of one-dimensional systems can be obtained from the relaxation of initial states featuring global (left-right) gradients in the chemical potential or temperature. The approach is tested for spinless interacting fermions as well as for the Fermi-Hubbard model, and the behaviour in the vicinity of special points (such as half filling or isotropic chains) is discussed. We present technical details on how to implement the calculation in practice using the density matrix renormalization group and show that the non-equilibrium dynamics is often less demanding to simulate numerically and features simpler finite-time transients than the corresponding linear response current correlators; thus, new parameter regimes can become accessible. As an application, we determine the thermal Drude weight of the Hubbard model for temperatures $T$ which are an order of magnitude smaller than those reached in the equilibrium approach. This allows us to demonstrate that at low $T$ and half filling, thermal transport is successively governed by spin excitations and described quantitatively by the Bethe ansatz Drude weight of the Heisenberg chain.

\end{abstract}


\maketitle

\section{Introduction}
\label{sec:intro}

Computing correlation effects on static or dynamic transport properties at finite temperature such as charge or thermal conductivities \cite{mahan,luttinger64} 
\begin{equation}\label{sigma}
\sigma_\nu(\omega) = 2\pi D_\nu\delta(\omega) + \sigma_\nu^\tn{reg}(\omega)
 \end{equation}
generally poses a daunting task for theorists. Even in linear response and at low energies (temperatures), the DC conductivity of gapless systems is usually not governed by the Luttinger liquid fixed point alone but influenced by the existence of conserved quantities \cite{Sirker2009}. In order to connect to actual experimental transport measurements on (quasi) 1D systems such as carbon nanotubes or strongly anisotropic 3D materials, it is thus essential to directly study microscopic Hamiltonians such as Heisenberg spin chains, spin ladders, or Hubbard models. This is a hard task even for seemingly `simple', Bethe-ansatz solvable systems (such as the Heisenberg spin chain), because -- similarly to correlation functions -- transport coefficients are determined by couplings between all excitations.

Whether or not a physical system exhibits dissipationless transport is signaled by the Drude weight $D_\nu$ in Eq.~(\ref{sigma}). For $D_\nu\neq0$, an initially excited current does not fully decay but will survive to infinite time. If for a given model the current operator $I$ is conserved by the Hamiltonian, $[I,H]=0$, transport is dissipationless at any temperature $T$. If $I$ does not commute with $H$ but has a finite overlap with (quasi-) local conserved quantities, dissipation processes are restricted and the Drude weight is non-zero; this can be shown strictly using the Mazur inequality \cite{suzuki71,mazur69,zotos97}.  While the question of dissipationless transport is mainly investigated for closed systems within linear response, it can also be studied in non-equilibrium setups \cite{langer09,jesenko11,langer11,karrasch14} or for open quantum systems \cite{prosen09,znidaric11,znidaric13a,mendoza-arenas13}.

One prototypical low-dimensional model is given by spinless, interacting fermions (equivalently, a Heisenberg XXZ spin chain); it can be diagonalized exactly using the Bethe ansatz \cite{essler-book,giamarchi} and possesses an infinite number of local conserved quantities. The energy current operator commutes with the Hamiltonian, and the corresponding Drude weight $D_\tn{th}$ was computed analytically \cite{kluemper02,sakai03}. The charge Drude weight $D_\tn{c}$ has attracted considerable attention \cite{zotos99,benz05,jung07,narozhny98,hm03,mukerjee08,alvarez02,heidarian07}, in particular at half filling where the current operator has no overlap with any of the standard local conserved quantities. Considerable progress has been made within the recent years \cite{Sirker2009,karrasch14,prosen11,herbrych11,karrasch13,prosen13,steinigeweg14,prelovsek04,steinigeweg09,sirker11,steinigeweg12}. In particular, Prosen constructed quasi-local conserved quantities \cite{prosen11,prosen13,mierzejewski14} to show analytically that $D_\tn{c}$ is finite throughout the gapless phase (excluding the isotropic point); quantitative numbers can be obtained, e.g., using the real-time density matrix renormalization group (DMRG) \cite{karrasch13,karrasch12} or dynamical typicality \cite{steinigeweg14}. Whether or not the Drude weight is finite for an isotropic chain is still debated \cite{herbrych11,karrasch13,steinigeweg14,carmelo15}.

A more complex (and more experimentally relevant) system is the 1d Fermi-Hubbard model, which is again integrable via the Bethe ansatz and can thus in principle support dissipationless transport at finite temperature; however, neither the charge nor the energy current are fully conserved, $[I_\tn{c,th},H]\neq0$. Most prior studies concentrated on the charge Drude weight $D_\tn{c}$ \cite{fujimoto98,kirchner99,peres00,carmelo12,prosen12,prosen14,karrasch14a,Jin2015}, which is finite away from half filling by virtue of the Mazur inequality \cite{zotos97}. Directly at half filling, most works point towards $D_\tn{c}(T)=0$ \cite{carmelo12,prosen12,karrasch14a,Jin2015}, but this issue is not fully resolved yet. The thermal Drude weight of the Hubbard model has attracted far less attention: While the Mazur inequality \cite{zotos97} can again be used to show that $D_\tn{th}(T)>0$ for arbitrary fillings, quantitative numbers for $D_\tn{th}$ were only computed recently for large-to-intermediate temperatures \cite{hubprl}. It is one goal of this work to obtain the thermal Drude weight via the DMRG for temperatures which are an order of magnitude smaller and to demonstrate that one successively recovers the exact form of the Heisenberg chain's thermal Drude weight at low $T$ and half filling. We can reach such small temperatures by extracting $D_\tn{th}$ using a novel numerical protocol (see the next paragraph) which differs from the standard one employed in Ref.~\cite{hubprl}.

The `standard route' to compute the Drude weight $D_\nu(T)$ numerically is provided by the linear response expression
\begin{equation}\label{eq1}
 D_\tn{c,th}(T) = \lim_{t\to\infty}\lim_{L\to\infty} \frac{\langle I_\tn{c,th}(t)I_\tn{c,th}\rangle_\tn{eq}}{2LT^{1,2}}\,,
\end{equation}
where the real-time current correlation function $\langle I_\nu(t)I_\nu\rangle_\tn{eq}$ can be obtained directly using the DMRG. It was recently demonstrated \cite{thermalpaper2} that $D_\nu(T)$ can alternatively be calculated from the non-equilibrium current $\langle I_\nu(t)\rangle_{\mu,T}$ flowing in the presence of a small chemical potential or temperature gradient via \cite{Ilievski2013}
\begin{equation}\label{noneq}
D_\tn{c,th}(T) = \lim_{t\to\infty}\lim_{L\to\infty} \partial_{\mu,T} \frac{\langle I_\tn{c,th}(t)\rangle_{\mu,T}} {2t}\,.
\end{equation}
Eq.~(\ref{noneq}) was discussed briefly in Ref.~\cite{thermalpaper2}, and its validity was tested explicitly for the XXZ spin chain. The aim of the present paper is to expand on the ideas of Ref.~\cite{thermalpaper2}, to study the practical relevance of Eq.~(\ref{noneq}) in more detail, and, as an application, to extract $D_\tn{th}$ of the Fermi-Hubbard model at low $T$.

After briefly introducing our methodology in Secs.~\ref{sec:model} and \ref{sec:dw}, we extensively compare the real-time dynamics of Eqs.~(\ref{eq1}) and (\ref{noneq}) in Sec.~\ref{sec:comp}. One particular focus is on charge and thermal transport in the Hubbard model (which was not considered in Ref.~\cite{thermalpaper2}). We discuss the behaviour in the vicinity of special points such as half filling or `isotropic chains'. In Sec.~\ref{sec:num}, practical aspects are presented on how to implement the calculation of Eq.~(\ref{noneq}) numerically. In particular, we document that Eq.~(\ref{noneq}) is often less demanding to simulate and features less complex finite-time transients than Eq.~(\ref{eq1}). As an application, we exploit this simplicity to determine the thermal Drude weight of the Hubbard model for temperatures which are an order of magnitude lower than those reached in the linear response calculation of Ref.~\cite{hubprl}, allowing us to access the regime where thermal transport is successively governed by spin excitations and described quantitatively by the exact $D_\tn{th}(T)$ of the Heisenberg chain (Sec.~\ref{sec:hubheis}).

\section{Model and Method}
\label{sec:model}

\subsection{Model}

The first model we consider describes spinless, interacting lattice fermions, whose Hamiltonian is given by
\begin{equation}\label{xxz}
H = \sum_l h_l = J \sum_l\left[ \frac{1}{2} \left(c^\dagger_l c^{\phantom{\dagger}}_{l+1} +\textnormal{h.c.}\right) + \Delta \tilde n_l\tilde n_{l+1}  \right]\,,
\end{equation}
with $c_l$ being a fermionic annihilation operator acting on site $l$, $\tilde n_l= c_l^\dagger c_l^{\phantom{\dagger}}-1/2$, and $J,\Delta$ denoting the hopping amplitude and nearest-neighbor interaction strength, respectively. Eq.~(\ref{xxz}) can be mapped to an XXZ spin chain with an exchange coupling $J$ and anisotropy $\Delta$ via a Jordan-Wigner transformation. The charge and energy current of this model take the standard form
\begin{equation}
 I_\tn{c} = \frac{iJ}{2}\sum_l \left( c_{l+1}^\dagger c_l^{\phantom{\dagger}}- \textnormal{h.c.}\right)\,,~~~
 I_\tn{th} = i\sum_l [h_l,h_{l+1}]\,.
\end{equation}
The one-dimensional Fermi-Hubbard model is governed by
\begin{equation}\begin{split}\label{hub}
H = \sum_l h_l = \sum_l\left[-t_0  \sum_\sigma \left(c_{l\sigma}^\dagger c_{l+1\sigma}^{\phantom{\dagger}} + \tn{h.c.} \right)
+ \frac{U}{2}\left(\tilde n_{l\uparrow}\tilde n_{l\downarrow}+\tilde n_{l+1\uparrow}\tilde n_{l+1\downarrow}\right)\right]\,,
\end{split}\end{equation}
where $c_{l\sigma}$ annihilates a fermion with spin $\sigma$ on site $l$, and $\tilde n_{l\sigma}=c_{l\sigma}^\dagger c_{l\sigma}^{\phantom{\dagger}}-1/2$. The interaction strength and the hopping matrix element are denoted by $U$ and $t_0$, respectively. The charge and energy current are given by
\begin{equation}\label{currentdef}
 I_\tn{c} = it_0\sum_l\sum_\sigma \left( c_{l+1\sigma}^\dagger c_{l\sigma}^{\phantom{\dagger}}- \textnormal{h.c.}\right)\,,~~~
 I_\tn{th} = i\sum_l [h_{l+1},h_{l}]\,.
\end{equation}

\subsection{Density Matrix Renormalization Group}

In order to calculate the real time evolution of the one-dimensional quantum-mechanical systems introduced in Eqs.~(\ref{xxz}) and (\ref{hub}), we employ the time-dependent \cite{vidal04,white04,daley04,schmitteckert04,vidal07} density matrix renormalization group method \cite{white92,schollwoeck05,schollwoeck11} implemented using matrix product states \cite{fannes91,ostlund91,verstraete06,verstraete08}. Finite temperatures \cite{Verstraete2004p,white09,barthel09,zwolak04,sirker05,barthel13} are incorporated via a purification $|\Psi\rangle_T$ of the thermal density matrix $\rho\sim e^{-H/T}$. The state $|\Psi\rangle_T$ can be obtained from the (known) $|\Psi\rangle_\infty$ via an evolution $e^{-H/2T}$ in the inverse temperature \cite{schollwoeck11}. Both $e^{-H/2T}$ as well as the real time evolution operator $e^{-iHt}$ are factorized by a fourth order Trotter-Suzuki decomposition. We keep the discarded weight during each individual `bond update' below a threshold value $\epsilon$. This leads to an exponential increase of the bond dimension $\chi$ during the real time evolution. In order to access time scales as large as possible, we employ the finite-temperature disentangler introduced in Ref.~\cite{Karrasch2012}, which exploits the fact that purification is not unique to slow down the growth of $\chi$. Our calculations are performed using a system size of the order of $L\sim O(100)$ sites. By comparing to other values of $L$, we have ensured that $L$ is large enough for the results to be effectively in the thermodynamic limit \cite{Karrasch2013}.

\section{Computation of the Drude weight}
\label{sec:dw}

\subsection{Motivation of the non-equilibrium expression for the linear Drude weight}

For reasons of completeness, we briefly \textit{motivate} the origin of Eq.~(\ref{noneq}) for the charge case -- more details can be found in Ref.~\cite{thermalpaper2}. Linear response theory predicts that local currents $i_\tn{c}(x)$ are related to gradients of an applied potential $\mu(x)$ via $i_{\tn{c}}(x)=-\sigma_\tn{c}\partial_x\mu(x)$. The spatially integrated current flowing in a large but finite system (see below for comments on this issue) is then given by
\begin{equation}\label{dwderiv}
\langle I_\tn{c}(t)\rangle_\mu = \int i_{\tn{c}}(x) dx = 2\pi D_\tn{c} \delta(\omega=0)\delta\mu + \ldots
 = 2 D_\tn{c} t \delta\mu +\ldots~,
\end{equation}
where $\delta\mu$ is the total potential difference, and we have exploited that finite times serve as an infrared cutoff and regularize the $\delta$-function via
\begin{equation}
\delta(\omega=0)\approx \int_{-t}^t \frac{dt}{2\pi} = \frac{t}{\pi}\,.
 \end{equation}
The ellipsis in Eq.~(\ref{dwderiv}) denotes a contribution from the regular part of the conductivity which can be neglected in the asymptotic limit of large times.

Hence, Eq.~(\ref{dwderiv}) suggests that the total non-equilibrium current flowing in the presence of an initial potential gradient should increase linearly for large times and that the prefactor is determined by the Drude weight. If the regular contribution to the conductivity vanishes and transport is purely ballistic, the finite-time transients should vanish and linear behaviour should manifest even for small $t$. We will explicitly verify this picture for the XXZ chain as well as for the Hubbard model.

One might wonder why for a fully ballistic system whose total current $I$ commutes with the Hamiltonian, $\langle I\rangle$ is not constant but increases linearly with time. This confusion can be resolved by recapitulating the meaning of boundary conditions. In Eq.~(\ref{dwderiv}), we have implicitly assumed that our system has open boundaries, that the potential gradients occur in its center, and that the system size $L$ is large enough so that at a time $t$ the perturbations spreading out from the center have not yet reached the boundaries (so that the system is practically in the thermodynamic limit). Put differently, the global current $I$ is effectively determined by the integral over a finite region, $\langle I\rangle=\int_{-a}^a i(x)dx$, whose size $a$ fulfills $vt\ll a\ll L$, with $v$ being the Lieb-Robinson velocity. Importantly, $[I,H]=0$ generally holds only for systems with periodic (not open) boundary conditions; the standard example is the energy current in the XXZ chain. If in our setup the left and right ends of the open system are connected, this creates a second, identical potential gradient, and the total current flowing in its vicinity is up to a minus sign identical to the one flowing in the center. Hence, the total current for a system with periodic boundary conditions is indeed constant. Note that this intuitive argument can be confirmed explicitly using the DMRG.

As a side remark, we note that if that Drude weight vanishes and transport is purely diffusive, the present setup contains information about the diffusion constant \cite{karrasch14} via the second moment of the spatial profile of the local currents in Eq.~(\ref{currentdef}). We leave a more thorough study for future work.

\subsection{Numerical details}

The linear response expression for the Drude weight, which is given by Eq.~(\ref{eq1}), can be simulated directly using the DMRG. It is advantageous to `exploit time translation invariance' \cite{barthel13},
\begin{equation}\label{eq2}
 D_\tn{c,th}(T) = \lim_{t\to\infty}\lim_{L\to\infty} \frac{\langle I_\tn{c,th}(t/2)I_\tn{c,th}(-t/2)\rangle_\tn{eq}}{2LT^{1,2}}\,,
\end{equation}
and to carry out two independent calculations for $I_\nu(t/2)$ as well as $I_\nu(-t/2)$. Combined with the finite-temperature disentangler \cite{Karrasch2012}, this allows one to reach time scales which are roughly four times as large as the ones accessible by a `standard' DMRG approach \cite{barthel09}. In principle, a similar `trick' can be implemented when calculating the out-of-equilibrium expression in Eq.~(\ref{noneq}) \cite{kennes16}; however, this is not necessary for our purposes.

In order to compute the Drude weight via Eq.~(\ref{noneq}), we initialize the system in a state
\begin{equation}
 \langle \ldots \rangle_{\mu,T} = \tn{Tr}\, \langle\rho_{\mu,T} \ldots \rangle_{\mu,T}
\end{equation}
featuring a gradient in the temperature or chemical potential. The latter case is straightforward: We simply prepare the system using a thermal density matrix
\begin{equation}
 \rho_\mu \sim e^{-\tilde H/T}\,,
\end{equation}
where $\tilde H$ is the Hamiltonian of Eq.~(\ref{xxz}) or Eq.~(\ref{hub}) complemented by a term $\pm J\frac{\delta \mu}{2}\tilde n_{l}$ for the XXZ chain [and similarly $\pm t_0\frac{\delta\mu}{2}(\tilde n_{l\uparrow}+\tilde n_{l\downarrow})$ for the Hubbard model] on sites $l\leq L/2$ and $l>L/2$, respectively. Furthermore, one can distinguish the cases in which the central bond connecting sites $L/2$ and $L/2+1$ is cut ($h_{L/2}$ set to zero) or not cut in $\tilde H$. The time evolution is then calculated using the original Hamiltonian (the potential is switched off).

The simplest way to compute the thermal Drude weight via Eq.~(\ref{noneq}) is to prepare the system in a state
\begin{equation}
 \rho_T \sim \rho_L\otimes\rho_R\,,
\end{equation}
where $\rho_{L,R}$ are thermal density matrices of separated left and right systems (sites $l\leq L/2$ and $l>L/2$), respectively. Their temperatures are chosen as
\begin{equation}
T_R=T\,,~ T_L=(1-\delta\beta)T_R\,.
\end{equation}
In this setup, the bond between the sites $L/2$ and $L/2+1$ is naturally cut. This can be circumvented (which will turn out to be advantageous numerically; see below) by preparing the system using a density matrix
\begin{equation}
 \rho_T \sim e^{-\tilde H/T}\,,
\end{equation}
where $\tilde H$ is the original Hamiltonian for sites $l>L/2$  and the original Hamiltonian multiplied by $1-\delta\beta$ for sites $l\leq L/2$, respectively. The real time evolution is again governed by the original $H$ given in Eq.~(\ref{xxz}) or Eq.~(\ref{hub}).

\begin{figure}
\begin{center}
\includegraphics[width=0.48\linewidth,clip]{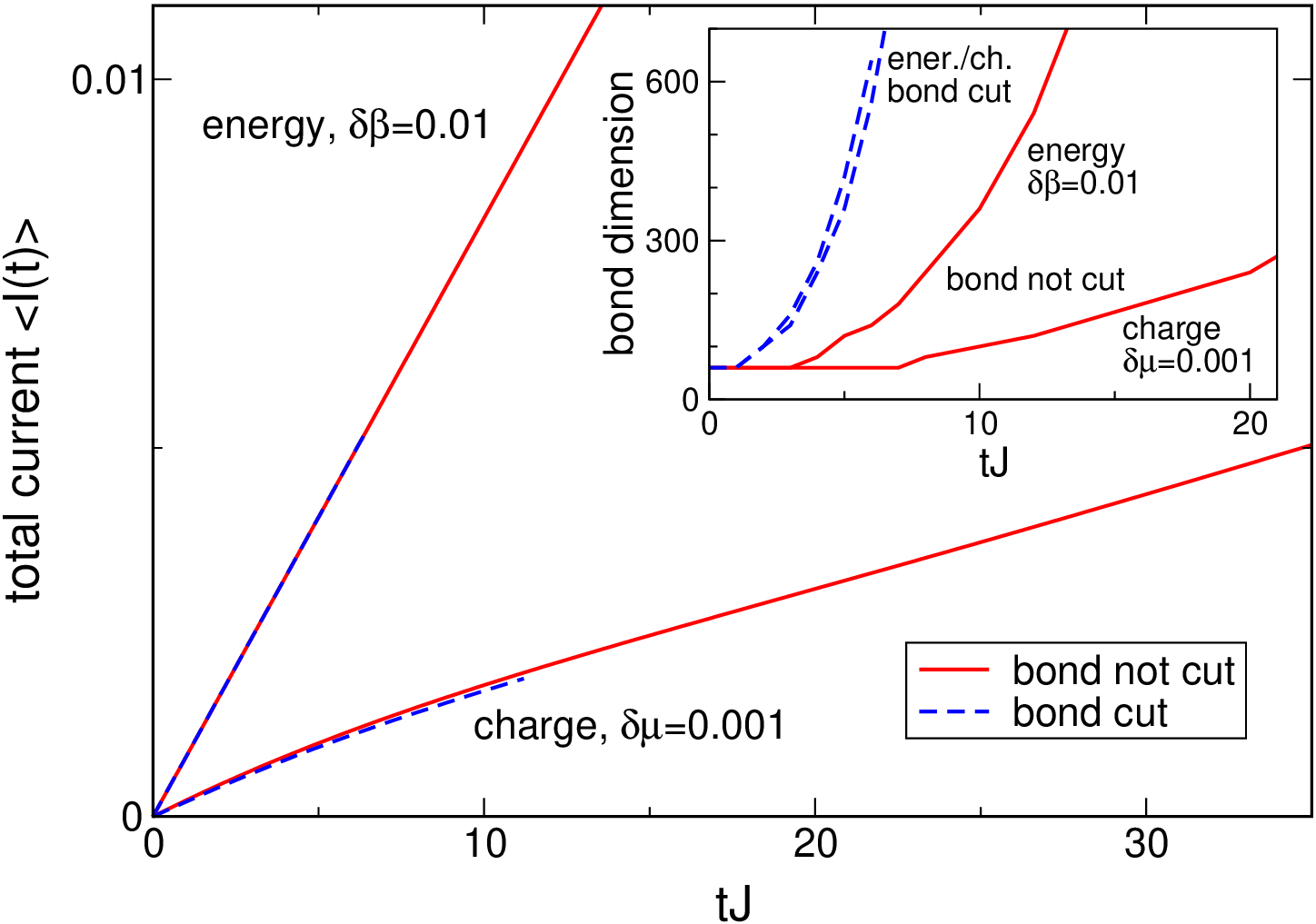}
\caption{(Color online) Total charge and energy currents $\langle I_\tn{c}\rangle/J$ and $\langle I_\tn{th}\rangle/J^2$ flowing in a half-filled chain of spinless lattice fermions exhibiting a nearest-neighbour interaction $\Delta=0.5$ and a temperature $T/J=0.5$ [see Eq.~(\ref{xxz})]. At time $t=0$, the chain is prepared in a state featuring a small, sharp gradient $\delta\mu$ or $\delta\beta$ in the chemical potential or temperature between the left and right halves; we explicitly compare the cases in which the bond connecting the two halves is cut (or not cut) in the preparation of this state [see Sec.~\ref{sec:dw} for details]. Inset: Evolution of the bond dimension during the different simulations.}
\label{fsetup}
\end{center}
\end{figure}

\section{Comparison with linear response}
\label{sec:comp}

In this section, we explicitly verify the validity of Eq.~(\ref{noneq}) for spinless fermions as well as for the Hubbard model and show that linear response Drude weights can indeed be obtained from the evolution of an out-of-equilibrium initial state featuring global gradients $\delta\mu$ or $\delta\beta$ in the chemical potential or temperature, respectively. We discuss the finite-time dynamics of Eqs.~(\ref{eq1}) and (\ref{noneq}) and demonstrate that they exhibit different decay rates as one approaches special points of vanishing Drude weights.

In Fig.~\ref{fsetup}, we show the time evolution of the total charge and energy currents $\langle I_\tn{c}(t)\rangle_\mu$ and $\langle I_\tn{th}(t)\rangle_T$ for spinless interacting fermions. While $[I_\tn{th},H]=0$ at any $\Delta$, $I_\tn{c}$ is not fully conserved, but the charge Drude weight is finite for $|\Delta|<1$ at any $T>0$ \cite{prosen11} and zero if $|\Delta|>1$. Hence, one expects that the non-equilibrium charge current grows linearly only for large times [since there is a non-vanishing regular contribution to the conductivity in Eq.~(\ref{sigma})] but that $\langle I_\tn{th}\rangle\sim t$ for all $t$ (see the discussion at the end of Sec.~\ref{sec:dw} to resolve a potential confusion related to the choice of boundary conditions). This is indeed the case for all parameters that we studied (and illustrated explicitly in Fig.~\ref{fsetup}; this Figure will be discussed in more detail in Sec.~\ref{sec:num}).

\begin{figure}
\begin{center}
\includegraphics[width=0.48\linewidth,clip]{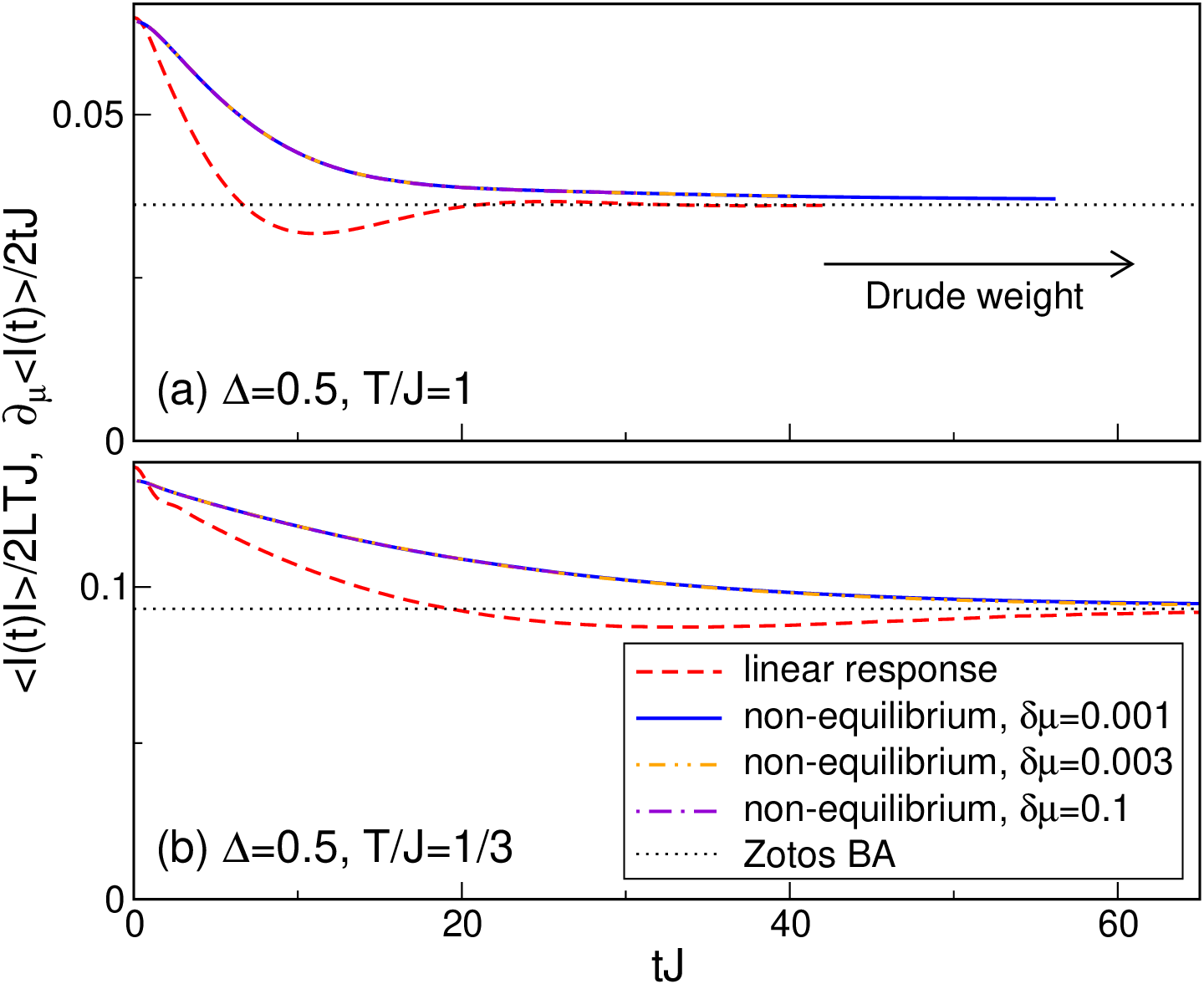}\hspace*{0.02\linewidth}
\includegraphics[width=0.48\linewidth,clip]{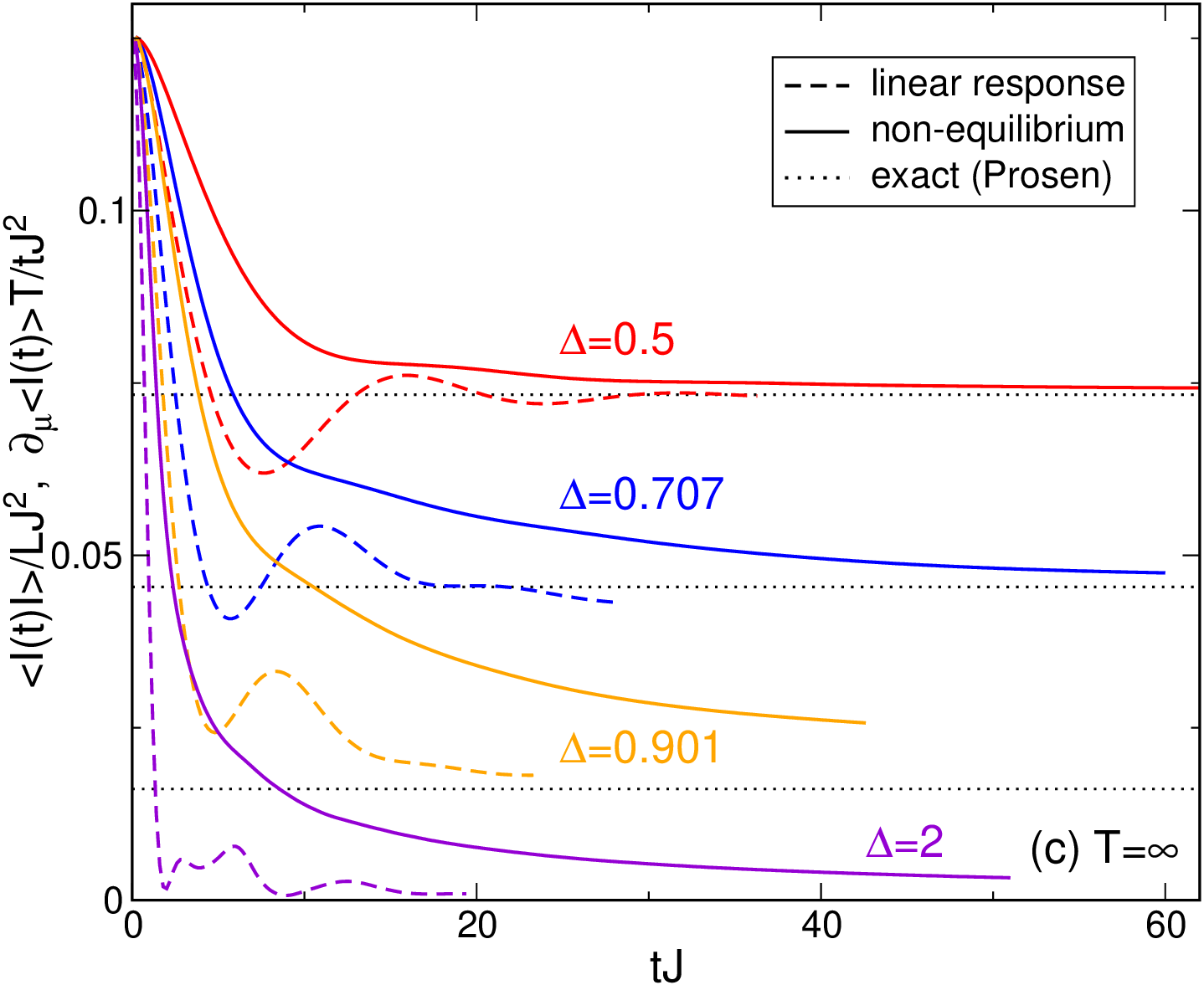}
\caption{(Color online) Real time evolution of the linear response charge current correlation function $\langle I_\tn{c}(t)I_\tn{c}\rangle_\tn{eq}$  as well as of the non-equilibrium current $\langle I_\tn{c}(t)\rangle_\mu$ induced by an initial gradient $\delta\mu$ in the chemical potential for a half-filled model of spinless fermions with a nearest-neighbour interaction $\Delta$. The long-time asymptote of both quantities yields the charge Drude weight by virtue of Eqs.~(\ref{eq1}) and (\ref{noneq}), respectively. At infinite temperature, the exact solution constructed by Prosen for $\Delta<1$ \cite{prosen11,prosen13} is shown as a reference; for finite $T$, we include Zotos' Bethe ansatz result from Ref.~\cite{zotos99}. The behaviour in the vicinity of the isotropic point $\Delta=1$ is discussed in the main text. } 
\label{fxxz}
\end{center}
\end{figure}

\subsection{Charge case}

We now explicitly compare the finite-time behaviour of the linear response charge current correlation function $\langle I_\tn{c}(t)I_\tn{c}\rangle_\tn{eq}$ and the non-equilibrium current $\langle I_\tn{c}(t)\rangle_\mu/t$ induced by an initial gradient $\delta\mu$ in the chemical potential. The large-$t$ asymptote of both quantities determines the Drude weight via Eqs.~(\ref{eq1}) and (\ref{noneq}), respectively.

Fig.~\ref{fxxz} shows data for spinless fermions at half filling and (a) $T/J=1$ at $\Delta=0.5$, (b) $T/J=1/3$ at $\Delta=0.5$, and (c) $\Delta\in\{0.5,0.707,0.901,2.0\}$ at $T=\infty$. In all cases, the long-time asymptotes of the linear response correlator and the out-of-equilibrium current agree. This confirms the validity of Eq.~(\ref{noneq}). At infinite temperature, an exact analytic solution for the charge Drude weight was constructed by Prosen \cite{prosen11,prosen13} and is shown as a reference in Fig.~\ref{fxxz}(c); see also Refs.~\cite{zotos99,benz05}. While the linear response correlators converge to the exact asymptote on a time scale which seems to be roughly independent of $\Delta$, the currents $\langle I_\tn{c}\rangle_\mu/t$ decay more slowly (towards zero for $\Delta>1$ or towards the Prosen bounds for $\Delta<1$) as one approaches the isotropic point $\Delta=1$ from either side. This is interesting since it is still debated whether or not $D_\tn{c}(T)$ is finite at $\Delta=1$ \cite{herbrych11,karrasch13,steinigeweg14,carmelo15}, and the out-of-equilibrium setup discussed in this paper might provide a new route to investigate this issue. If one fits the data at large times to an exponential function, one observes that the corresponding decay rate increases as one approaches $\Delta=1$. However, at the same time the quality of the fit worsens since the time scale accessible by the DMRG decreases [compare the curves at $\Delta=0.707$ and $\Delta=0.901$ in Fig.~\ref{fxxz}(c)]. Analysing this more quantitatively is not straightforward and left for future work.

\begin{figure}
\begin{center}
\includegraphics[width=0.48\linewidth,clip]{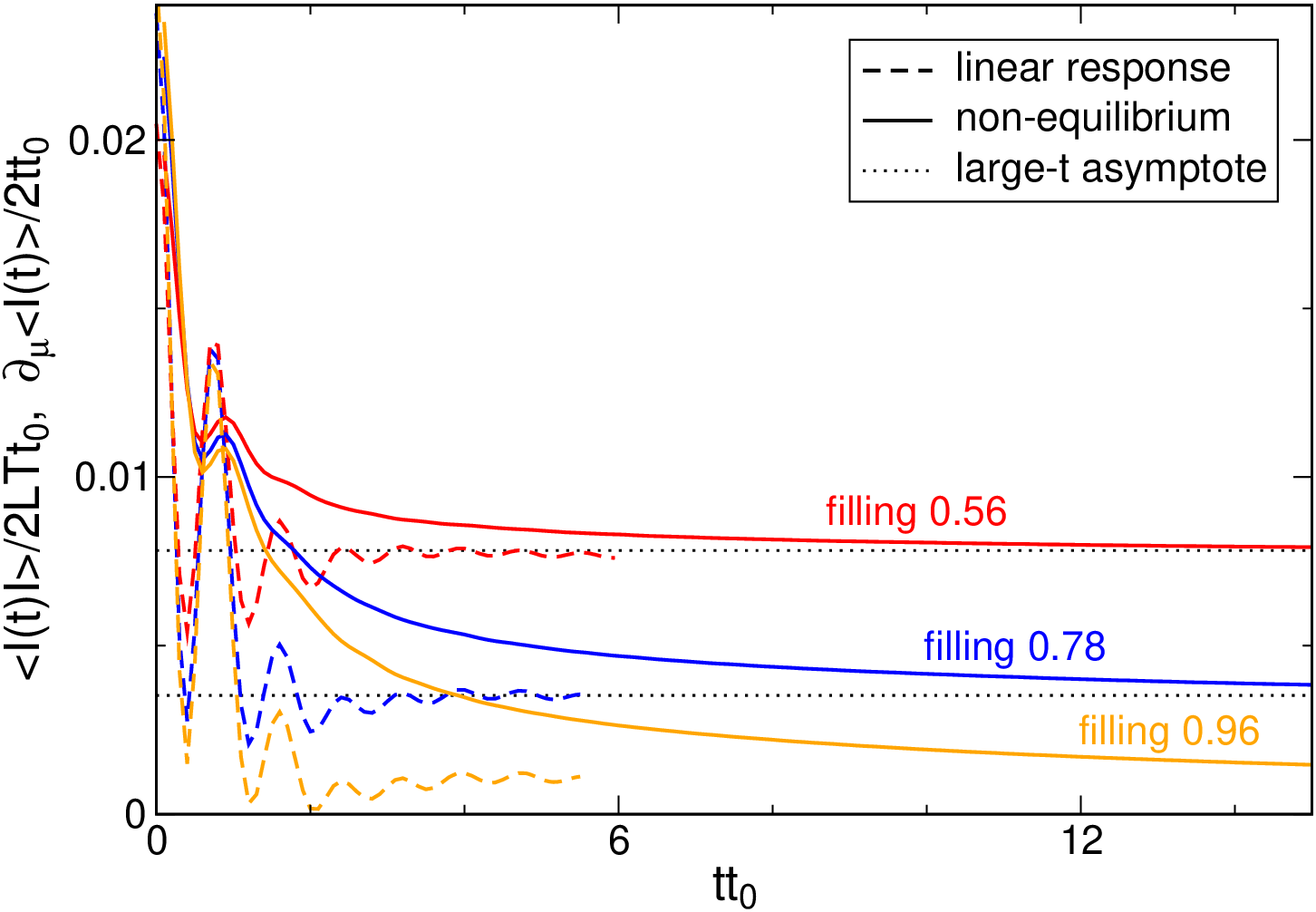}
\caption{(Color online) The same as in Fig.~\ref{fxxz}, but for charge currents in the Fermi-Hubbard model [see Eq.~(\ref{hub})] with an on-site interaction $U/t_0=8$, temperature $T/t_0=20$, and three filling factors. The large-time asymptote was determined by an exponential fit of the non-equilibrium data for $tt_0\gtrsim 8$ and fillings $0.56,0.78$.}
\label{fhub}
\end{center}
\end{figure}

Next, we study charge transport in the Fermi-Hubbard model. While the Drude weight is finite away from half filling by virtue of the Mazur inequality \cite{zotos97}, most works point towards $D_\tn{c}(T)=0$ directly at half filling \cite{carmelo12,prosen12,karrasch14a,Jin2015}, but this issue is not finally resolved. Fig.~\ref{fhub} shows the linear response correlators as well as the non-equilibrium currents for an on-site interaction of strength $U/t_0=8$ at a temperature of $T/t_0=20$ for three values of the filling. Both quantities converge to the same asymptotic value, which again validates Eq.~(\ref{fhub}). Moreover, we observe that the currents $\langle I_\tn{c}\rangle_\mu/t$ follow a simple exponential decay at large times, and sufficiently away from half filling one can more reliably determine $D_\tn{c}$ by fitting to this form (see the dotted lines in Fig.~\ref{fhub}). Interestingly, it seems that while the non-equilibrium currents decay more slowly as one approaches half filling, the linear response correlators do not exhibit a similar qualitative change but level off on comparable time scales. This scenario is analogous to what happens in the vicinity of $\Delta=1$ for spinless fermions and might again be used to gain further insights about the -- still not fully resolved -- issue of the charge Drude weight at half filling (future work).

\subsection{Thermal case}

Next, we turn to the thermal Drude weight. For spinless fermions, the energy current operator commutes with the Hamiltonian -- transport is always purely ballistic. This is no longer the case in the Hubbard model, but the Mazur inequality can be used to prove that $D_\tn{th}(T)>0$ for arbitrary fillings \cite{zotos97}. Hence, no subtleties occur at special points (in contrast to the charge case), and the asymptotic behaviour of $\langle I_\tn{th}(t)I_\tn{th}\rangle_\tn{eq}$ and $\langle I_\tn{th}(t)\rangle_T/t$ can be determined straightforwardly. This is illustrated for two sets of parameters in Fig.~\ref{fhubth}(a,b). We therefore do not present real-time data in more detail but directly discuss results for the Drude weight obtained via Eqs.~(\ref{eq1}) and (\ref{noneq}), respectively.

Fig.~\ref{fth} shows linear response and out-of-equilibrium data for $D_\tn{th}$ as a function of the temperature for (a) spinless fermions with $\Delta\in\{0.5,1\}$, and (b) the Hubbard model with $U/t_0\in\{0,4,8\}$, both at half filling. In (a), we plot the exact Bethe ansatz solution \cite{kluemper02} for comparison; note that in (b), the point $U/t_0=0$ can be solved analytically, and we show the exact linear response result instead of the DMRG data. The high-$T$ asymptote (dashed lines) displays $\langle I_\tn{th}(t)I_\tn{th}\rangle_{\tn{eq},T=\infty}/2LT^2$. In both models and for all temperatures and interactions, the Drude weight extracted using Eq.~(\ref{noneq}) agrees with the linear response prediction. This again confirms the validity of the non-equilibrium approach.

\subsection{Final thoughts}

\begin{figure}
\begin{center}
\includegraphics[width=0.48\linewidth,clip]{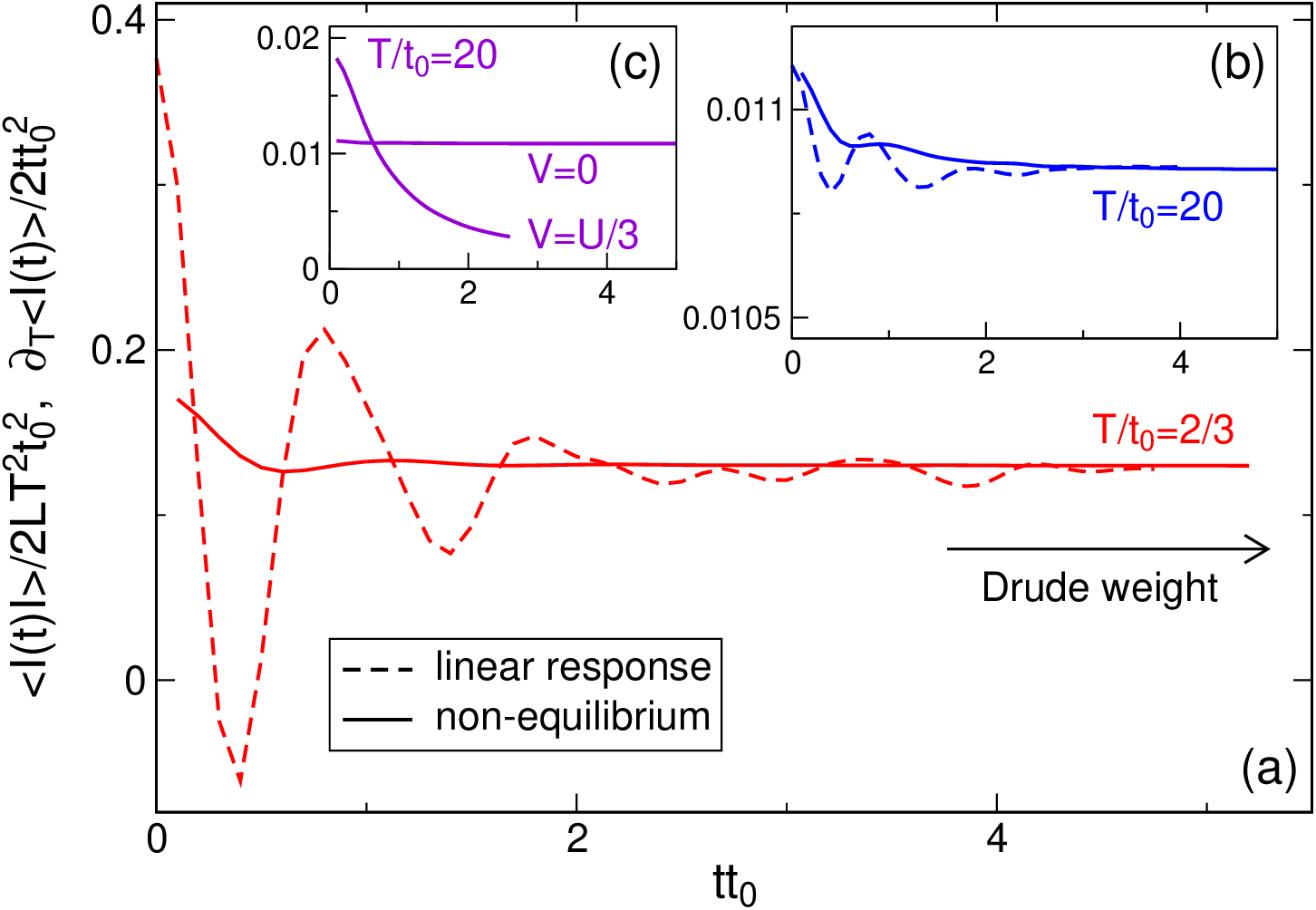}
\caption{(Color online) Linear response energy current correlation function as well as the non-equilibrium energy current induced by an initial small temperature gradient for the half-filled Fermi-Hubbard model with $U/t_0=8$ and (a) $T/t_0=2/3$, (b) $T/t_0=20$. The long-time asymptotes determine the thermal Drude weights via Eqs.~(\ref{eq1}) and (\ref{noneq}), respectively. In (c), we show data in the presence of an additional integrability-breaking nearest-neighbour interaction $V$.}
\label{fhubth}
\end{center}
\end{figure}

If the integrability of the model at hand is broken, charge and thermal Drude weights become zero, and the non-equilibrium currents $\langle I_\tn{c,th}(t)\rangle_{\mu,T}/t$ decay to zero. We have verified this explicitly for charge and thermal transport in the Hubbard model in presence of an additional nearest-neighbor interaction $V$; representative results are presented in Fig.~\ref{fhubth}(c).

To summarize, we have shown that charge and thermal Drude weights can be obtained either from the linear response correlators using Eq.~(\ref{eq1}) or from out-of-equilibrium currents via Eq.~(\ref{noneq}). While both expressions yield the same asymptotic value, the finite-time transients do not necessarily agree. This becomes particularly obvious as one approaches special points of potentially vanishing Drude weights. Pragmatically, the non-equilibrium currents often exhibit a simpler (e.g., non-oscillatory) transient behaviour [see, e.g., Fig.~\ref{fhubth}(a)], which renders it simpler to extract the Drude weight away from those special points.

\section{Computational details}
\label{sec:num}

In this Section, we present data for different initial states and illustrate how small the gradients in the chemical potential or temperature need to be chosen in practice in order to recover the linear response prediction. Moreover, we compare the numerical effort necessary to simulate Eqs.~(\ref{eq1}) and (\ref{noneq}), respectively.

Fig.~\ref{fsetup} shows the charge and energy currents $\langle I_\tn{c}\rangle_\mu$ and $\langle I_\tn{th}\rangle_T$ for spinless fermions and two different initial states. The bond connecting the left and right regions (between which the initial gradients $\delta\mu$ and $\delta\beta$ occur) is cut in one of them by setting $h_{L/2}=0$ but left unchanged in the other (see Sec.~\ref{sec:dw} for details on how the state is actually prepared). The currents feature the same asymptotic behaviour in both cases, and even the finite-time transients (which appear in the charge case) are small. However, the numerical effort is drastically reduced if the bond is not cut in the preparation of the state (see the inset to Fig.~\ref{fsetup}), which one can understand intuitively from the fact that by setting $h_{L/2}=0$, one chooses an initial state which is further away from the stationary one. Hence, it is numerically advantageous to not `cut the bond' in the preparation of the initial state, and all data in this work was obtained using this setup.

\begin{figure}
\begin{center}
\includegraphics[width=0.48\linewidth,clip]{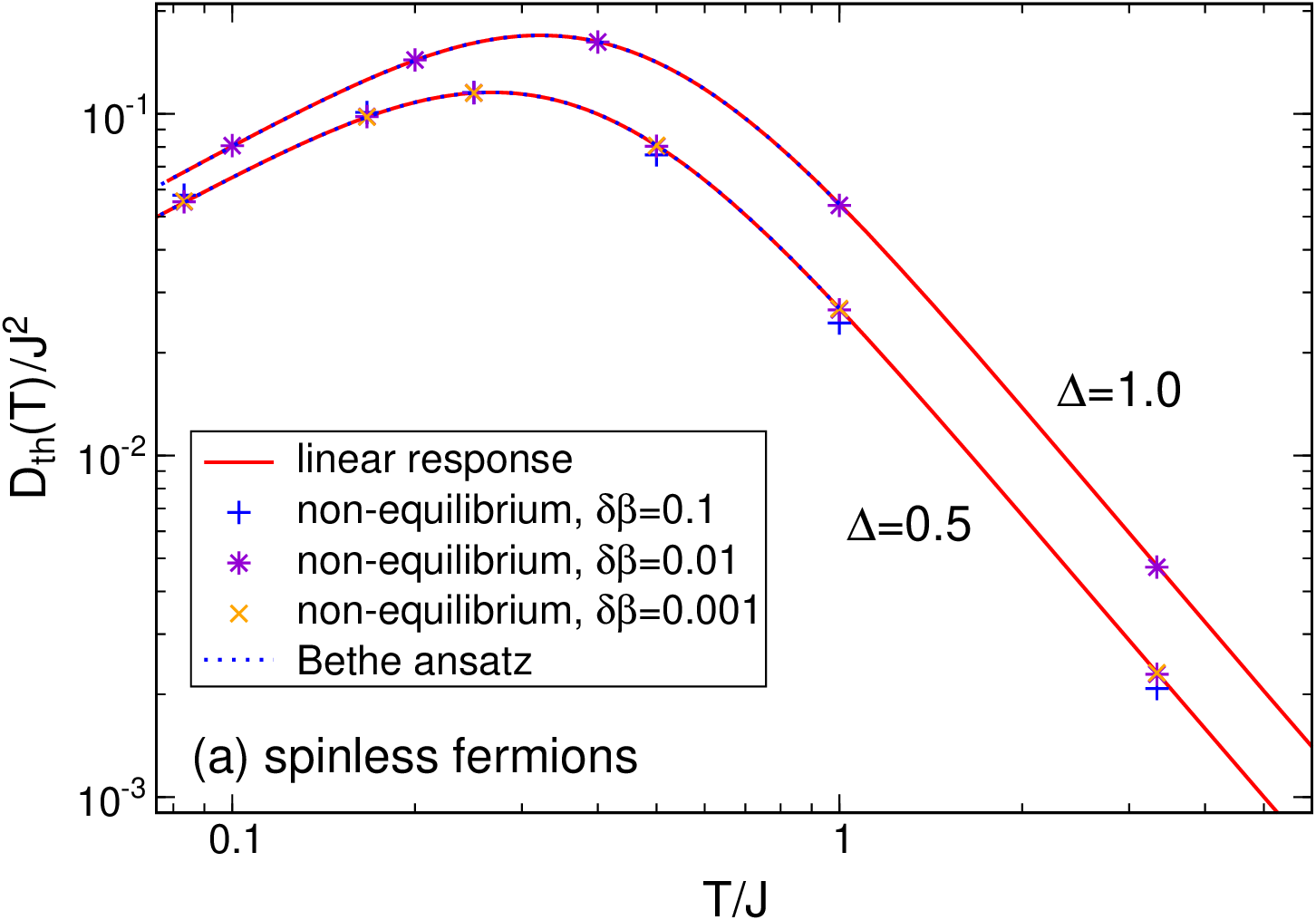}\hspace*{0.02\linewidth}
\includegraphics[width=0.48\linewidth,clip]{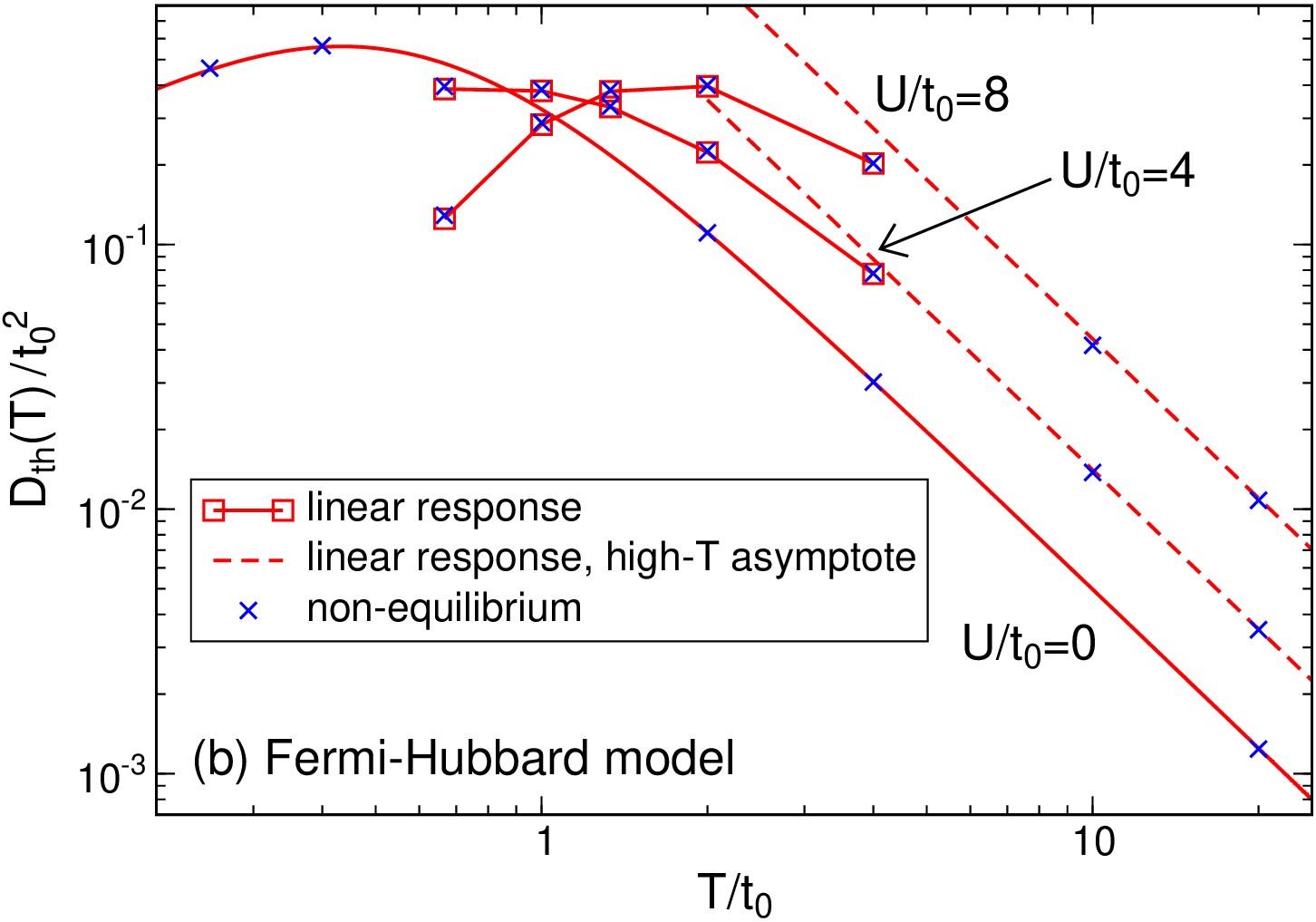}
\caption{(Color online) Thermal Drude weight $D_\tn{th}(T)$ for (a) spinless fermions and (b) the Fermi-Hubbard model at half filling. The linear response result calculated using Eq.~(\ref{eq1}) is compared to the one obtained via Eq.~(\ref{noneq}) from the non-equilibrium energy current $\langle I_\tn{th}(t)\rangle_T$ induced by an initial temperature gradient $\delta\beta$. In (a), we show the exact Bethe ansatz result \cite{kluemper02} for $T/J\leq 1$ as a reference (the Bethe ansatz and linear response DMRG curves are indistinguishable). In (b), the linear response Drude weight was computed analytically for $U/t_0=0$.}
\label{fth}
\end{center}
\end{figure}

In Figs.~\ref{fxxz}(a) and \ref{fth}(a), we explicitly show non-equilibrium data for spinless fermions calculated for different strength $\delta\mu$ and $\delta\beta$ of the initial potential and temperature gradients. One can see that $\delta\mu,\delta\beta\sim0.01$ is small enough to reproduce the linear response result with an accuracy that is beyond the resolution of the Figure; deviations only occur for $\delta\beta=0.1$ in Fig.~\ref{fth}(a). All other data in this work was obtained using $\delta\mu,\delta\beta\sim0.01-0.1$, and we checked (in representative cases) that decreasing the gradients even further does not influence the results.

It is instructive to recall that the order of limits $\delta\mu\to0$ and $t\to\infty$ in Eq.~(\ref{noneq}) is defined on an operational level: One first prepares a gradient $\delta\mu$ and then time-evolves until the DMRG breaks down (at a finite time scale). This procedure is repeated with successively decreasing $\delta\mu$ (starting from fairly large $\delta\mu$) until the results (on the accessible time scales) no longer change. This is illustrated in Fig.~\ref{fxxz} for $\delta\mu\in\{0.1,0.003,0.001\}$.

Applying the real- and imaginary time evolution operators $e^{-iHt}$ and $e^{-H/T}$ to a matrix product state involves singular value decompositions which lead to an increase of the bond dimension. The key approximation of the DMRG is to truncate this state by discarding the singular values below of a given threshold. The allowed discarded weight is the central parameter which controls the accuracy of the method.

\begin{figure}
\begin{center}
\includegraphics[width=0.48\linewidth,clip]{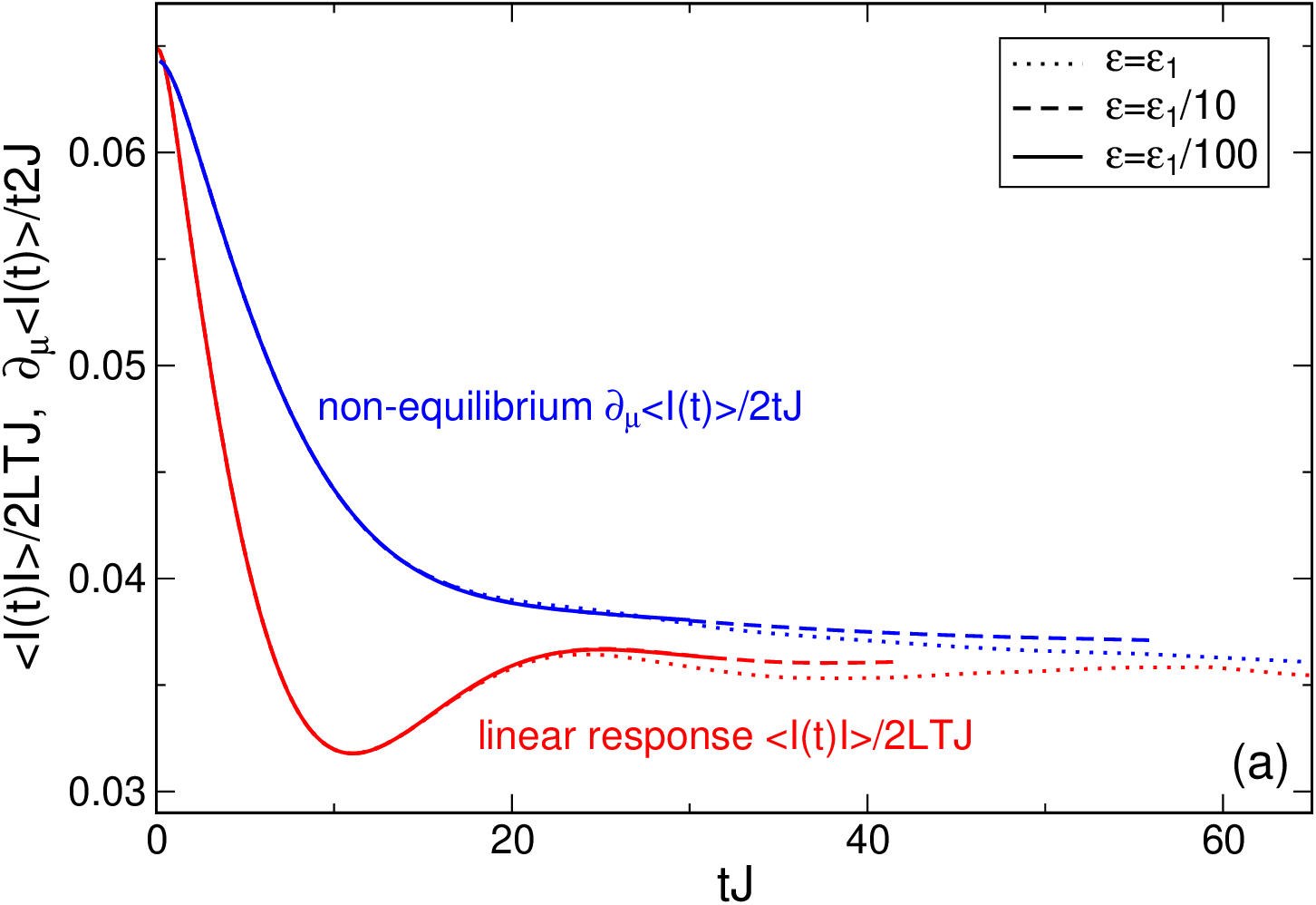}\hspace*{0.02\linewidth}
\includegraphics[width=0.48\linewidth,clip]{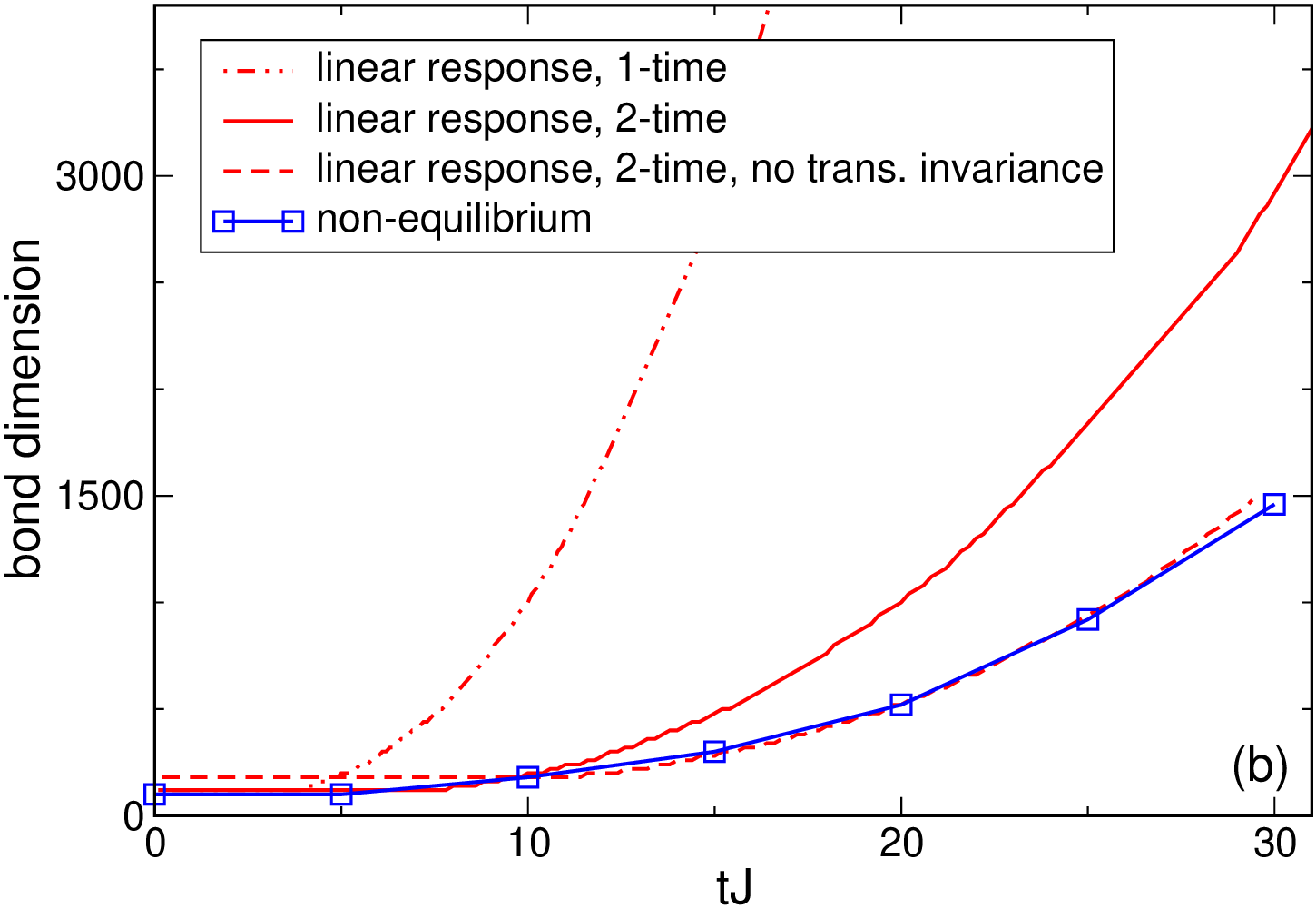}
\caption{(Color online) (a) The same as in Fig.~\ref{fxxz}(a) but for three different values of the discarded weight $\epsilon$, which in total varies over two orders of magnitude (the non-equilibrium result is for $\delta\mu=0.001$). (b) Evolution of the bond dimension during the calculation of the data for the smallest $\epsilon$. The meaning of the different linear-response curves is explained in the main text.}
\label{fnum} 
\end{center}
\end{figure}

In practice, we choose some representative sets of physical parameters and carry out calculations using different values of the discarded weight $\epsilon$ during the real time evolution. An example is shown in Fig.~\ref{fnum}(a), which displays the data of Fig.~\ref{fxxz}(a) for three different, successively decreasing $\epsilon\in\{\epsilon_1,\epsilon_1/10,\epsilon_1/100\}$: We start from a large $\epsilon_1$ and then lower this value succesively until the physical quantity at hand is computed with the desired accuracy. Note that (i) the bond dimension grows faster for smaller $\epsilon_1$, and hence the accessible time scales are reduced, and (ii) the linear response and non-equilibrium calculations are generally performed using a different $\epsilon_1$ chosen such that the corresponding curves $\langle I_\nu(t)I_\nu\rangle_\tn{eq}$ and $\langle I_\nu(t)\rangle_{\mu,T}$ eventually reach the desired accuracy. In this work, the desired accuracy is set by the scale of each plot: In the case of Fig.~\ref{fnum}(a), no deviation between the data calculated for $\epsilon=\epsilon_1/10$ and $\epsilon=\epsilon_1/100$ can be observed [on the scale of Fig.~\ref{fnum}(a)]; hence, the former value is a reasonable choice.

In Fig.~\ref{fnum}(b), we illustrate how the bond dimension $\chi$ grows if the smallest value $\epsilon_1/100$ is chosen as the discarded weight. We compare $\chi(t)$ for the simulations of (i) the linear response expression $\langle I_\tn{c}(t)I_\tn{c}\rangle_\tn{eq}$ calculated in the standard way from a single time evolution, (ii) the same, but writing $\langle I_\tn{c}(t)I_\tn{c}\rangle_\tn{eq}=\langle I_\tn{c}(t/2)I_\tn{c}(-t/2)\rangle_\tn{eq}$ and carrying out two individual time evolutions for $I_\tn{c}(\pm t/2)$, and (iii) the non-equilibrium approach $\langle I_\tn{c}(t)\rangle_\mu$. The fastest growth occurs in (i). Using (ii), one can access a time scale which is (roughly) twice as large at the same computational cost. More precisely, translation invariance (in space) can only be exploited in one of the $I_\tn{c}(\pm t/2)$; their calculations thus exhibit different $\chi(t)$ (and are also performed using different individual discarded weights $\epsilon_{1}>\epsilon_{1,\tn{no\,trans.\,inv.}}$; see Ref.~\cite{kennes16} for details). If translation invariance is exploited, the bond dimension $\chi$ at a time $t$ is identical to $\chi(t/2)$ of the standard, single-time approach [1-time and 2-time curves in Fig.~\ref{fnum}(b)]; it still grows significantly faster than in the non-equilibrium approach. If translation invariance is not exploited in the linear response simulation, the growth of the bond dimension is comparable to the one of the non-equilibrium calculation. However, the former simulation is much more demanding, especially at low temperatures (we postpone arguments to the next paragraph). Hence, one can conclude that for this set of parameters the non-equilibrium calculation is the least computationally challenging one and can therefore be performed up to larger times. From a purely pragmatic standpoint, one should note that in order to obtain $\langle I_\nu(t)\rangle_{\mu,T}$, one simply needs to time-evolve a state which is determined by the purification of the initial, non-equilibrium density matrix. In contrast, the linear response approach in its two-time version requires the calculation of a correlation function $\langle I_\nu(t/2)I_\nu(-t/2)\rangle_\tn{eq}$, which is more difficult to implement numerically. Extracting Drude weights via Eq.~(\ref{noneq}) thus seems to be a viable alternative to the standard linear response route.

We conclude this Section with a few more technical remarks; additional details can again be found in Ref.~\cite{kennes16}. If one does or does not exploit translation invariance in the linear response approach, the calculation of $I_\nu(\pm t/2)$ amounts to time-evolving locally or globally quenched states $I_{\nu,L/2}|\Psi\rangle_T$ or $I_{\nu}|\Psi\rangle_T$, respectively. The non-equilibrium setup always corresponds to time-evolving a locally quenched state. In local quenches, perturbations spread with a finite Lieb-Robinson velocity, and the bond dimension does not increase significantly outside of this `light cone'. This is one reason why the linear response calculation is more demanding if one cannot use translation invariance. Moreover, one needs to perform the time evolution of $I_{\nu}|\Psi\rangle_T$, which requires the application of a global operator $I_\nu$ to the state $|\Psi\rangle_T$. This increases the bond dimension instantaneously by a factor which is determined by the matrix product operator representation of $I_\nu$; all additional symmetries (such as spin-flip symmetry) should hence be exploited in this simulation and not in $I_{\nu,L/2}|\Psi\rangle_T$. Since finite temperatures are reached via an evolution in $1/T$ starting from $T=\infty$, $\chi$ grows with decreasing $T$. In practice, for the Hubbard model at moderately low $T$, $\chi$ can reach values of $\chi\sim1000$, and applying the global energy current operator $I_\tn{th}$ to $|\Psi\rangle_T$ and subsequently computing its time evolution becomes no longer feasible.

\begin{figure}
\begin{center}
\includegraphics[width=0.48\linewidth,clip]{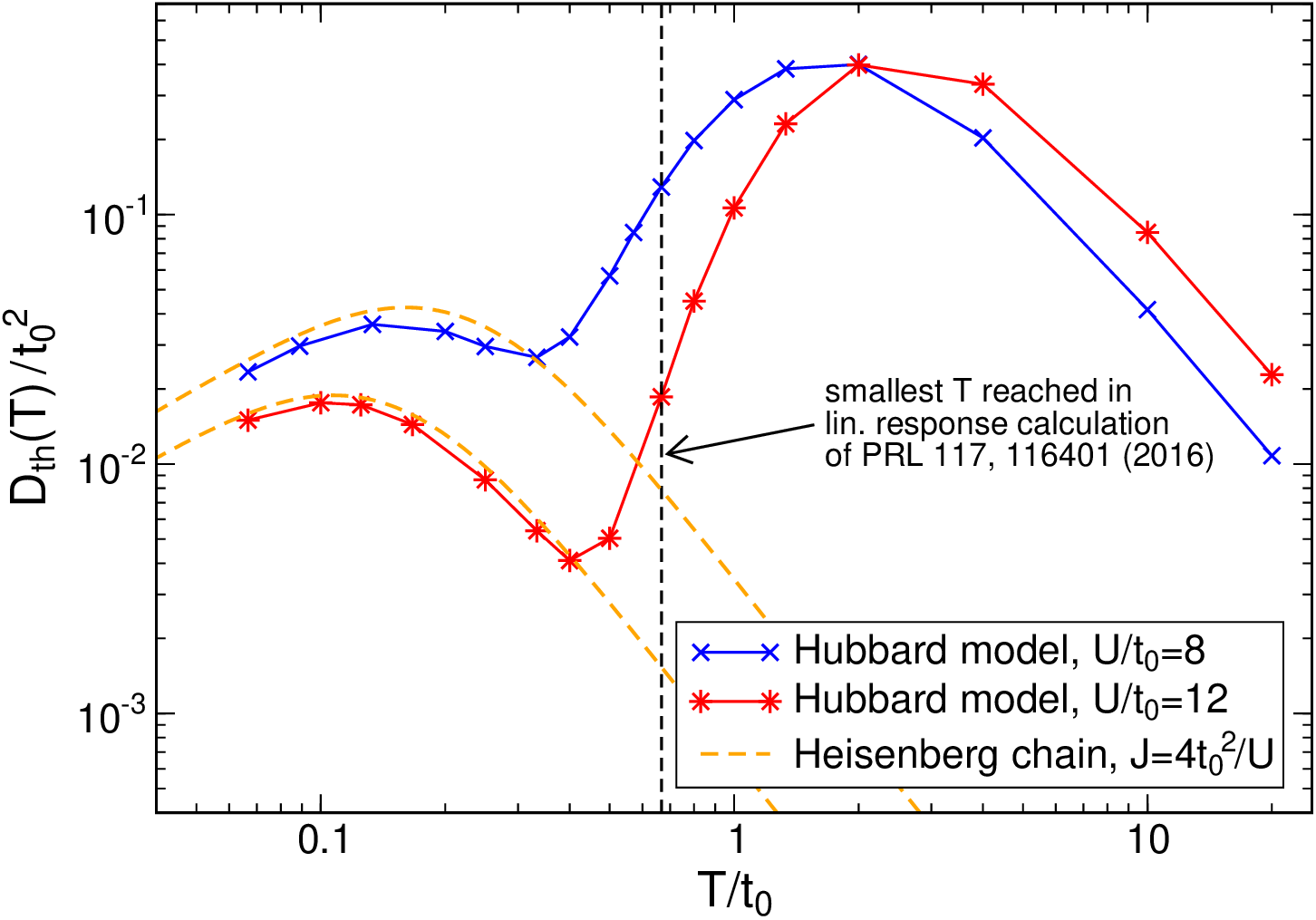}
\caption{(Color online) Thermal Drude weight $D_\tn{th}(T)$ of the half-filled Fermi-Hubbard model for two different values of the on-site interaction $U$. By extracting $D_\tn{th}$ using the non-equilibrium expression of Eq.~(\ref{noneq}), one can access temperatures which are an order of magnitude smaller than those reached in the linear response calculation \cite{hubprl}. The Figure demonstrates that at low $T$, thermal transport is governed by spin excitations and described quantitatively by the Bethe ansatz Drude weight of an isotropic XXZ spin chain with an exchange coupling $J=4t_0^2/U$.} 
\label{fhubheis}
\end{center}
\end{figure}

\section{Hubbard model: thermal Drude weight at low $T$}
\label{sec:hubheis}

In this Section, we revisit the realm of the thermal Drude weight of the Fermi-Hubbard model. As mentioned above, the Mazur inequality \cite{zotos97} stipulates that $D_\tn{th}(T)$ is finite at any filling, and quantitative values were recently obtained \cite{hubprl} using the linear response expression of Eq.~(\ref{eq1}). In Fig.~\ref{fth}(b), we explicitly demonstrated that the Drude weight extracted from the time evolution of an initial, small temperature gradient via Eq.~(\ref{noneq}) coincides with the linear response prediction.

We can now exploit the computational simplicity of the non-equilibrium approach as well as the fact its finite-time transients have a simpler form (see Fig.~\ref{fhubth}) to determine the thermal Drude weight for temperatures which are an order of magnitude lower than those reached in Ref.~\cite{hubprl}. The results are shown in Fig.~\ref{fhubheis} at half filling and for two values $U/t_0\in\{8,12\}$ of the on-site interaction. At high temperatures, $D_\tn{th}$ increases quadratically, becomes maximal when $T$ reaches the charge gap, and subsequently decreases since charge excitations are frozen out. One expects that at low $T$, transport is governed by spin excitations whose dynamics are described by a Heisenberg spin chain [or equivalently, Eq.~(\ref{xxz})] with an exchange coupling of strength $J=4t_0^2/U$. In other words, one expects to recover a second peak in $D_\tn{th}(T)$ at low $T$ whose form quantitatively follows the exact (Bethe ansatz or DMRG) Drude weight of the Heisenberg chain [the curve at $\Delta=1$ in Fig.~\ref{fth}(a) with units rescaled to $J=4t_0^2/U$]. This is indeed the case. To the best of our knowledge, this two-peak structure constitutes the first quantitative observation of a full Hubbard-to-Heisenberg crossover for a transport quantity within the one-dimensional Fermi-Hubbard model at finite temperature.

\section{Summary}

In this paper, we have investigated how the linear response charge and thermal Drude weights of integrable one-dimensional systems can be computed from the relaxation of initial states featuring small gradients in the chemical potential or temperature. Using density matrix renormalization group numerics for spinless fermions as well as for the Hubbard model, we extensively compared the real-time dynamics of the currents $\langle I_\tn{c,th}(t)\rangle_{\mu,T}/t$ flowing in this non-equilibrium setup with the linear response correlators $\langle I_\tn{c,th}(t)I_\tn{c,th}\rangle_\tn{eq}$. While both quantities determine $D_\tn{c,th}(T)$ in the limit $t\to\infty$, the finite-time behaviour differs. Only $\langle I_\tn{c,th}(t)\rangle_{\mu,T}/t$ seems to exhibit diverging decay rates in the vicinity of points where it is still debated whether or not the Drude weight vanishes; we explicitly demonstrated this for charge transport in an XXZ spin chain near $\Delta=1$ by comparing with Prosen's exact solution. Away from such special points, the non-equilibrium currents often exhibit simpler (e.g., non-oscillatory) transients and are less demanding to compute numerically. We exploited this to extract the thermal Drude weight of the Hubbard model for temperatures which are an order of magnitude lower than those reached in the linear response approach. At half filling and sufficiently large on-site interactions, $D_\tn{th}(T)$ features a two-peak structure and at low $T$ is quantitatively described by the exact Bethe ansatz Drude weight of the Heisenberg spin chain.

It would be interesting to generalize our approach in order to efficiently extract transport properties beyond the Drude weight (such as the low-frequency behaviour of the regular part of the conductivity). Moreover, the vicinity of special points (e.g., isotropic XXZ chains) certainly deserves further attention.

\section*{Acknowledgments}
I thank F.~Heidrich-Meisner and D.~M.~Kennes for discussions and acknowledge support by the DFG through the Emmy Noether program (KA 3360/2-1).

\vspace*{1cm}

\bibliographystyle{iopart-num}
\bibliography{references}

\end{document}